\documentclass[useAMS,usenatbib]{mn2e}
\usepackage{psfig}

\def\eiso{\mbox{$E_{\rm \gamma,iso}$}}
\def\fl{\mbox{$S_{\rm \gamma}$}}
\def\ep{\mbox{$E_{\rm pk}$}}
\def\epr{\mbox{$E_{\rm pk}^\prime$}}
\def\ebr{\mbox{$E_{0}$}}
\def\nor{\mbox{$N_0$}}
\def\slope{\mbox{$\alpha$}}
\def\tdur{\mbox{T$_{\rm int}$}}
\def\sp{\mbox{$\nu f_\nu$~}}

\def\sw{\mbox{\it Swift}} 
\def\hete{\mbox{HETE--2}}

 \def\spose#1{\hbox to 0pt{#1\hss}}
 \newcommand\lsim{\mathrel{\spose{\lower 3pt\hbox{$\mathchar"218$}}
      \raise 2.0pt\hbox{$\mathchar"13C$}}}
 \newcommand\gsim{\mathrel{\spose{\lower 3pt\hbox{$\mathchar"218$}}
      \raise 2.0pt\hbox{$\mathchar"13E$}}}

\title[Spectral analysis of Swift GRBs with known $z$]
      {Spectral analysis of Swift long GRBs with known redshift}

\author[Cabrera et al.]
{J.~I. Cabrera$^1$\thanks{E--mail: jcabrera@astroscu.unam.mx},
C. Firmani$^{1,2}$, V. Avila--Reese$^1$, G. Ghirlanda$^2$, G. Ghisellini$^2$ and 
\newauthor 
L. Nava$^{2,3}$\\
$^{1}$Instituto de Astronom\'ia, Universidad Nacional Aut\'onoma de M\'exico, 
       A.P. 70--264, 04510, M\'exico, D.F.\\
$^{2}$Osservatorio Astronomico di Brera, via E.Bianchi 46, I--23807 Merate, Italy\\
$^{3}$Universit\'a degli studi dell'Insubria, 
Dipartimento di Fisica e Matematica, via Valleggio 11, I--22100 Como, Italy}

\begin{document}



\maketitle


\begin{abstract}
We study the spectral and energetics properties of 47
long--duration gamma--ray bursts (GRBs) with known redshift, all of them
detected by the \sw\ satellite. Due to the narrow energy range (15--150 keV) 
of the \sw--BAT detector, the spectral fitting is reliable 
only for fitting models with 2 or 3 parameters. As high uncertainty and correlation 
among the errors is expected, a careful analysis of the errors is necessary.
We fit both the power law (PL, 2 parameters) and cut--off power law (CPL, 3
parameters) models to the time--integrated spectra of the 47 bursts,
and we present the corresponding parameters, their uncertainties, and
the correlations among the uncertainties.  
The CPL model is reliable only for 29 bursts for which we estimate the \sp\
peak energy \ep. For these GRBs, we calculate the energy fluence
and the rest--frame isotropic--equivalent radiated energy, \eiso, as well as the 
propagated uncertainties and correlations among them.  We explore the distribution 
of our {\it homogeneous} sample of GRBs on the rest--frame diagram \epr\ vs 
\eiso. We confirm a significant correlation between these two quantities 
(the ``Amati" relation) and we verify that, within the uncertainty limits, no 
outliers are present. We also fit the spectra to a Band model with the 
high energy power law index frozen to $-2.3$, obtaining a rather good agreement 
with the ``Amati" relation of non--\sw\ GRBs.

\end{abstract}


\section{Introduction}

The \sw\ Mission \citep{Gehrels} was designed mainly to 
rapidly detect, locate, and observe gamma--ray bursts (GRBs).
After more than two years of operation, the \sw\ satellite
has observed approximately 180 GRBs, for which more than 50
events have known redshifts reported in the Mission 
homepage\footnote{http://swift.gsfc.nasa.gov/}.
In this paper we present results 
from an {\it homogeneous} spectral analysis that we have carried out 
to the \sw\ long--duration GRBs with known $z$ with the main aim of estimating
the time--averaged spectral parameters of the prompt emission of these bursts, 
as well as related global quantities like the prompt energy fluence, \fl, and
the rest--frame isotropic--equivalent radiated energy, \eiso.

Unfortunately, the spectral coverage of the \sw\ $\gamma$--ray detector,
the Burst Alert Telescope \citep[BAT,][]{Barthelmy}, is very narrow. 
The recommendation is to limit the spectral 
analysis to channels between 15 keV and 195 keV, though the \sw--BAT's 
team suggests a more conservative upper limit of 150 keV because,
above this energy, the calibration is not yet sufficiently reliable.
Thus, the BAT energy range is not broad enough to allow the spectra to be fitted 
unambiguously by the usual GRB broken power--law function of 4 parameters.
Therefore, the inference of $\gamma$--ray spectral parameters and global
quantities related to these parameters for the \sw\ GRBs is not an
easy task.

A main concern of this task is the determination of the uncertainties of the 
spectral parameters and their propagation into the composite quantities. 
Due to the narrow energy range of \sw~BAT we know {\it a priori} that the errors 
in the fitted spectral parameters will be large and correlated among them. 
Therefore, the 
appropriate handling of errors is crucial in order to make useful the spectral 
data generated by BAT. The effort is justified by the great value that 
an {\it homogeneous} sample has, where the spectral information for all the events 
is obtained with the same detector and analyzed with the same techniques and 
methods. This is the case of the growing--in--number sample of \sw\ GRBs.

The characterization of the time--averaged photon spectra of GRBs with known
redshift is important for several reasons. On one hand, this characterization
offers clues for understanding the radiation and particle acceleration mechanisms 
at work during the prompt phase of the bursts 
\citep[see for recent reviews e.g.,][]{ZM04, Piran05, Meszaros06, Zhang07}. 
On the other hand, the spectral parameters and the global quantities inferred 
from them (\fl, \eiso, etc.) are among the main properties that 
characterize GRBs. The study of the rest--frame correlations among these and 
other properties \citep[e.g,.][]{Amati02, Atteia, Ghirla04, yonetoku04, liang05, 
Firmani06} is currently allowing to learn key aspects of the nature of GRBs 
\citep[e.g.,][]{Thompson05, Thompson06,RR06}. Furthermore, those correlations that are 
tight enough can be used to standardize the energetics of GRBs, making possible 
to apply GRBs for constructing the Hubble diagram up to unprecedentedly high redshifts 
\citep[Ghirlanda et al. 2004b, 2006; Dai et al. 2004; Firmani et al. 
2005, 2006b, 2007;][]{liang05, XDL05, schaefer07}. These tight correlations, for 
a given  
cosmological model, can be used also to estimate the (pseudo)redshifts of GRBs  
with not measured redshifts, mainly from the extensive {\it CGRO}--BATSE database\footnote{
In a recent paper, \citet{Li07}  showed that due both to a redshift degeneracy in the so--called Amati 
correlation  \citep{Amati02} and to its high scatter, is not possible to infer reliable (pseudo)redshifts
from this correlation.}
\citep[e.g.,][]{Lloyd02, Atteia}. Thus, by combining the distribution in redshift of a 
large sample of GRBs with the observed flux distribution, inferences on the 
luminosity function and formation rate of GRBs can be obtained 
\citep{Firmani04, Guetta05}.

The outline of this paper is as follows. In \S 2 we explain the main
steps of the spectral analysis carried out by us: selection of the sample
(\S\S 2.1), spectral deconvolution (fit) procedure and fitting models (\S\S 2.2),
and error analysis (\S\S 2.3). The results of the 
spectral analysis are given in \S 3, along with the estimates
of \fl~ and \eiso~ as well as some correlations among the obtained
quantities. The summary and conclusions of the paper are presented in \S 4.  
In the Appendix the deconvolved spectra of the 47 GRBs studied here are plotted.

We adopt the concordance $\Lambda$CDM cosmology with $\Omega_M=0.3$,  
$\Omega_{\Lambda}=0.7$, and $H_0=70$ km s$^{-1}$Mpc$^{-1}$.

\section{Spectral analysis}
\label{analysis}

The first step in our spectral analysis is the selection of the 
\sw~ sub-sample of GRBs to be studied (\S\S 2.1). Then we proceed
to the spectral deconvolution with XSPEC and determination of
the GRB spectral parameters for  two-- and three--parameter photon models 
(\S\S 2.2). By using these parameters, the energy fluence and isotropic--equivalent 
energy are calculated. Finally, we construct the error confidence ellipses for 
the spectral parameters and \eiso~ (\S\S 2.3).

 The BAT is a large aperture $\gamma-$ray telescope. It consists of a 
(2.4$\times$1.2)m coded aperture mask supported one meter above the 5200 cm$^{2}$ 
area 
detector plane.  The detector contains 32,768 individual cadmium-zinc-telluride 
detector elements, each one of (4$\times$4$\times$2)mm size. The effective efficiency 
of BAT starts at approximately 15 keV, attains a broad maximum at approximately 
30--100 keV and then falls off, attaining at 195 keV the same level as at 15 keV. 
Due to the 1 mm thickness of the lead tiles, the BAT coded mask starts to be 
transparent around 150 keV. We reduce the BAT's data using the analysis techniques 
provided by the BAT's instrument 
team\footnote{http://heasarc.gsfc.nasa.gov/docs/swift/analysis/threads}.

\subsection{The sample}
\label{sample}
 
We analyse the time--integrated spectra of all available \sw~ GRBs with known 
redshift $z$. The list of events was taken from the daily updated compilation
of J. Greiner\footnote{http://www.mpe.mpg.de/~jcg/grbgen.html} for the period January 
2005 to January 2007. The corresponding file spectra were loaded from the
legacy NASA database\footnote{ftp://legacy.gsfc.nasa.gov/swift/data/obs}. The total 
number of events in this list was 55. We exclude from the list the short--duration
bursts ($T_{90}<2$s; GRB050509, GRB050724, and GRB061217) as well as those events 
with incomplete observational information (GRB050730, GRB060124, GRB060218, 
GRB060505 and GRB060512), leaving us with a preliminary list of 47 long GRBs with 
measured $z$.

The list of 47 events is given in Table \ref{PLfit}. The first four columns of 
Table \ref{PLfit} give respectively the burst name, the class (see below), the 
redshift $z$, and the time \tdur that we use to integrate the spectrum.
In principle T$_{90}$ could be used as an integration time. Nevertheless
two problems arise in this case. The first one has to do with the systematic 
underestimate of the fluence, hence of \eiso. The second problem may be more 
important; the elimination of the (early and late) wings in the light curve 
can influence the estimates of the spectral parameters. For example the elimination 
of the late wing may harden artificially the spectrum.
For these reasons we have fixed the integration time by a visual
inspection of the light curve as the period where the signal is
clearly identifiable in the background noise.
A modest overestimate of \tdur~ does not influence on the values of the
integrated parameters, though increases the noise.
We would like to remark that \tdur~ is not a good estimate of the burst duration
because its uncertainty is related to the identification of faint
(some time extended) wings.

\begin{table*} 
\begin{tabular}{lcccccllc} 
\hline   
GRB      & Class  &  $z$  &  $\tdur$  &  Lg\nor [sd]  & \slope [sd]  &  
$\hspace{0.11in} e_1 \hspace{0.25in} {\bf n_1}$ & $\hspace{0.11in} e_2 \hspace{0.25in} {\bf n_2}$ & $\chi^2_{\rm red}$   \\
         &        &       & sec  &  50 keV&              &       &        \\ 
\hspace{0.08in} (1) & (2) & (3) & (4) & (5) & (6) & \hspace{0.25in} (7) & \hspace{0.25in} (8) & (9) \\
\hline   
050126 & A & 1.29 & 40.0 & -2.656[0.026]& -1.317[0.101]& 0.026(-1.00, 0.01) & 0.106(-0.01,-1.00) & 1.399 \\
050223 & A & 0.5915 & 30.0 & -2.652[0.028]& -1.837[0.106]& 0.025(-0.99, 0.13) & 0.109(-0.13,-0.99) & 0.902 \\
050315 & D & 1.949 & 129.0 & -2.567[0.013] & -2.101[0.055] & 0.012(-0.99, 0.11) & 0.056(-0.11,-0.99) & 0.889\\
050318 & A & 1.44 & 34.5 & -2.375[0.014] & -1.968[0.057] & 0.012(-0.99, 0.13) & 0.060(-0.13,-0.99) & 0.936 \\
050319 & D & 3.24 & 190.0 & -3.100[0.034]& -2.045[0.118]& 0.026(-0.98, 0.17) & 0.119(-0.17,-0.98) & 0.687 \\
050401 & B & 2.9 & 43.0 & -1.667[0.012] & -1.472[0.046]& 0.011(-1.00,-0.04) & 0.047( 0.04,-1.00) & 1.060\\
050416 & D & 0.6535 & 5.0 & -2.225[0.055]& -3.119[0.155] & 0.026(-0.95, 0.30) & 0.182(-0.30,-0.95) & 0.971 \\
050505 & A & 4.27 & 70.0 & -2.418[0.018] & -1.415[0.071] & 0.018(-1.00,-0.03) &0.073( 0.03,-1.00) & 0.789 \\
050525 & A & 0.606 & 15.0 & -0.932[0.003] & -1.769[0.015] & 0.003(-1.00,-0.09) & 0.015( 0.09,-1.00) & 2.934 \\
050603 & B & 2.821 & 16.0 & -1.347[0.010] & -1.159[0.043] & 0.010(-1.00,-0.05) & 0.043( 0.05,-1.00) & 0.913 \\
050803 & A & 0.422 & 165.0 & -2.868[0.022] & -1.526[0.085]& 0.021(-1.00, 0.04) & 0.088(-0.04,-1.00) & 0.882 \\
050814 & A & 5.3 & 95.0 & -2.750[0.030]& -1.798[0.108]& 0.027(-0.99, 0.12) & 0.117(-0.12,-0.99) & 1.036 \\
050820 & C & 2.61 & 32.0 & -2.461[0.016] & -1.689[0.067] & 0.015(-1.00, 0.09) & 0.067(-0.09,-1.00) & 0.988 \\
050824 & D & 0.83 & 38.0 & -3.102[0.081]& -2.743[0.219] & 0.036(-0.95, 0.30) & 0.241(-0.30,-0.95) & 1.0157 \\
050904 & C & 6.29 & 207.0 & -2.603[0.009] & -1.205[0.041]& 0.009(-1.00,-0.04) & 0.041( 0.04,-1.00) & 1.070 \\
050908 & A & 3.344 & 24.0 & -2.685[0.031]& -1.947[0.107]& 0.025(-0.99, 0.17) & 0.114(-0.17,-0.99) & 0.969 \\
050922C & B & 2.198 & 10.0 & -1.761[0.009]& -1.371[0.038]& 0.009(-1.00,-0.02) & 0.038( 0.02,-1.00) & 0.900 \\
051016B & D & 0.9364 & 6.0 & -2.509[0.048] & -2.331[0.145]& 0.030(-0.97, 0.24) & 0.158(-0.24,-0.97) & 1.344 \\
051109 & A & 2.346 & 43.0 & -2.052[0.033] & -1.626[0.124] & 0.032(-1.00, 0.05) & 0.129(-0.05,-1.00) & 0.819 \\
051111 & B & 1.55 & 45.0 & -2.080[0.008]& -1.305[0.034]& 0.008(-1.00,-0.04) & 0.035( 0.04,-1.00) & 0.764 \\
051221 & C & 0.547 & 3.1 & -1.422[0.008] & -1.397[0.037]& 0.008(-1.00,-0.03) & 0.037( 0.03,-1.00) & 0.996 \\
060115 & A & 3.53 & 128.0 & -2.851[0.018]& -1.771[0.071]& 0.016(-0.99, 0.11) & 0.073(-0.11,-0.99) & 0.875 \\
060206 & A & 4.048 & 14.0 & -2.184[0.011]& -1.693[0.047]& 0.011(-1.00, 0.06) & 0.049(-0.06,-1.00) & 1.061\\
060210 & A & 3.91 & 278.0 & -2.579[0.015]& -1.544[0.059] & 0.015(-1.00,-0.01) & 0.060( 0.01,-1.00) & 0.928 \\
060223 & A & 4.41 & 14.8 & -2.320[0.020]& -1.775[0.078] & 0.019(-0.99, 0.11) & 0.082(-0.11,-0.99) & 1.009 \\
060418 & C & 1.489 & 80.0 & -1.993[0.006]& -1.634[0.030]& 0.006(-1.00,-0.01) & 0.030( 0.01,-1.00) & 0.815 \\
060502A & B & 1.51 & 34.0 & -2.157[0.011]& -1.431[0.045]& 0.011(-1.00,-0.01) & 0.046( 0.01,-1.00) & 1.031 \\
060510B & A & 4.9 & 289.0 & -2.839[0.011]& -1.778[0.048] & 0.010(-1.00, 0.07) & 0.049(-0.07,-1.00) & 0.643 \\
060522 & A & 5.11 & 80.0 & -2.829[0.024] & -1.578[0.093]& 0.023(-1.00, 0.08) & 0.098(-0.08,-1.00) & 1.188 \\
060526 & C & 3.21 & 320 & -3.387[0.035]& -1.924[0.140] & 0.028(-0.99, 0.17) & 0.132(-0.17,-0.99) & 0.967 \\
060604 & D & 2.68 & 67.0 & -3.356[0.085]& -2.040[0.283]& 0.060(-0.98, 0.21) & 0.290(-0.21,-0.98) & 0.804 \\
060605 & A & 3.711 & 37.0 & -2.810[0.026]& -1.474[0.105]& 0.025(-1.00, 0.07) & 0.109(-0.07,-1.00) & 0.912\\
060607 & A & 3.082 & 127.0 & -2.686[0.011] & -1.436[0.046] & 0.010( 1.00, 0.02) & 0.047(-0.02,-1.00) & 0.947 \\
060614 & C & 0.125 & 154.0 & -1.877[0.005] & -1.936[0.024]& 0.004(-1.00, 0.01) & 0.024( 0.01,-1.00) & 1.072 \\
060707 & A & 3.425 & 81.0 & -2.652[0.022] & -1.667[0.082]& 0.021(-1.00, 0.06) & 0.087(-0.06,-1.00) & 1.116 \\
060714 & A & 2.71 & 135.0 & -2.614[0.016] & -1.935[0.063]& 0.013(-0.99, 0.12) & 0.064(-0.12, 0.99) & 1.161 \\
060729 & C & 0.54 & 134.0 & -2.658[0.021]& -1.870[0.083]& 0.019(-0.99, 0.12) & 0.084(-0.12,-0.99) & 0.863 \\
060904B & A & 0.703 & 187.0 & -3.059[0.022] & -1.628[0.090]& 0.021(-1.00, 0.10) & 0.090(-0.10,-1.00) & 0.855 \\
060906 & A & 3.686 & 63.0 & -2.386[0.020] & -2.006[0.075] & 0.017(-0.99, 0.14) & 0.078(-0.14,-0.99) & 1.114\\
060908 & A & 2.43 & 30.0 & -2.040[0.009]& -1.339[0.039]& 0.009(-1.00,-0.01) & 0.040( 0.01,-1.00) & 0.864 \\
060926 & D & 3.2 & 13.0 & -2.724[0.044] & -2.460[0.146] & 0.026(-0.97, 0.24) & 0.151(-0.24,-0.97) & 0.972\\
060927 & A & 5.6 & 32.0 & -2.388[0.013] & -1.654[0.052]& 0.012(-1.00, 0.07) & 0.055(-0.07,-1.00) & 1.158 \\
061007 & C & 1.261 & 85.0 & -1.334[0.004]& -1.002[0.016] & 0.003(-0.99,-0.13) & 0.016( 0.13,-0.99) & 0.523 \\
061110B & C & 3.44 & 147.0 & -3.055[0.021]& -1.069[0.091] & 0.021(-1.00,-0.05) & 0.091( 0.05,-1.00) & 1.175 \\
061121 & C & 1.314 & 125.0 & -1.936[0.004] & -1.410[0.018]& 0.004(-1.00,-0.08) & 0.018( 0.08,-1.00) & 0.450 \\
061222B & A & 3.355 & 57.0 & -2.367[0.023] & -1.962[0.084]& 0.020(-0.99, 0.14) & 0.089(-0.14,-0.99) & 1.280 \\
070110 & C & 2.352 & 103.0 & -2.806[0.016]& -1.581[0.702]& 0.016(-1.00, 0.06) & 0.070(-0.06,-1.00) & 1.007 \\
\hline
\end{tabular} 
\caption{The sample of \sw\ long GRBs with known $z$ analysed here (47 events). 
The $z$ and duration (\tdur) of the burst as well as the results from the 
PL photon model fit are reported. Second column refers to the class group assigned to the 
burst according to our spectral analysis (see text). The best fit values of the parameters 
and their one--dimensional standard deviations (sd) are given in columns 5 and 6; $e_{1,2}$ 
and {\bf $n_{1,2}$} in columns 7 and 8 are the semi-axis lengths and principal axes of the 
error ellipse, respectively. The reduced $\chi^2$ is reported in the last column (dof=58). }  
\label{PLfit}
\end{table*}


\subsection{Spectral fits}

It is well known that the GRB photon spectra 
are in general well described, in the energy range of
$\sim 10$ keV to a few MeV, by the so--called Band function \citep{Band93}, 
which is a two smoothly connected power laws:
\begin{eqnarray}
 N(E)=N_{0}\left( \frac{E}{\rm 100~ keV}\right) ^{\alpha}\exp\left( -\frac{E}{E_{0}}\right), \nonumber  \\
 E \leqslant E_{0}(\alpha-\beta)\\
 = N_{0}\left[ \frac{E_{0}(\alpha-\beta)}{\rm 100~keV}\right] ^{(\alpha-\beta)} 
\exp(\beta-\alpha)\left( \frac{E}{\rm 100~keV}\right) ^{\beta}, \nonumber \\
 E  >  E_{0}(\alpha-\beta)    \nonumber
\label{band}
\end{eqnarray}
where $\alpha$ and $\beta$ are  the photon indices of the low and high 
energy power laws, respectively, and \ebr~ is the $e-$folding (break) 
energy related to the peak energy in the \sp [or $E^2f(E)$] spectrum by 
$\ep=\ebr\times(2+\alpha)$. Note that \ep~ is well defined for
$\alpha \ge -2$ and $\beta < -2$. The normalization $N_0$
is in photons s$^{-1}$ cm$^{-2}$ keV$^{-1}$.

The spectral range of BAT is narrower than that one of BATSE
and other previous detectors. In the most recent update of the BAT--team 
homepage\footnote{http://swift.gsfc.nasa.gov/docs/swift/analysis/bat-digest.html}
it is written that due to varying threshold levels in individual detectors, 
channels below 15 keV should not be used for spectral analysis. Likewise, 
channels above 150 keV are unreliable due to a lack of calibration data at 
those energies. Then,  we are limited to the narrow energy band of 15--150 
keV for the spectral analysis of \sw~ bursts. This implies that model 
fitting with the 4--parameter Band function will not be reliable for 
most of these burst. Therefore, we proceed to fit the observed photon 
spectra to both the power law (PL) and cutoff power law  (CPL) models. 
The former has only two parameters, the normalization, \nor,
and the photon index, \slope, and is given by:
\begin{equation}
 N(E)=N_{0}\left( \frac{E}{\rm 50~keV}\right)  ^{\slope}
\label{PL}
\end{equation}
The CPL model implies one more parameter, an energy $e-$folding spectral 
break, \ebr, related to \ep~ in the same way as in the Band model (see above).
The CPL model is defined as:
\begin{equation}
 N(E)=N_{0}\left( \frac{E}{\rm 50~keV}\right) ^{\slope}
\exp\left[  -\frac{E}{\ebr}\right]
\label{CPL}
\end{equation}
It is easy to verify that the CPL model (named also the Compton --COMP--
model) is the Band model with $\beta\rightarrow -\infty$.

As we will see later, the complex nature of the errors due to the small energy 
range of the spectra  makes it more convenient to use the logarithms of \nor~ and  
\ebr~ (or \ep) rather than the linear quantities. This choice reduces the asymmetry 
of the errors and of the confidence levels (CL) contour shapes. Therefore, we 
have introduced into XSPEC new models corresponding to Eqs. (\ref{PL}) and 
(\ref{CPL}) (and also 
Eq. (1); see \S\S 3.3) to carry out the fits with Log\nor~  and Log\ep.

The spectral analysis was carried out by using the heasoft6.1.2 public 
software\footnote{http://heasarc.gsfc.nasa.gov/docs/software/lheasoft/download.html}.
The first step in our analysis is to define the background--subtracted 
light curve from the corresponding data file. 
The photon counts at each channel are taken in the time interval \tdur\ where 
the main light curve is clearly above the background noise. The spectral file 
was corrected by position and systematic errors. Then, the photon counts at each
channel is convolved with a response matrix, build up for a given event with the 
calibration matrix\footnote{We use the January 2007 matrix calibration 
provided by the \sw--BAT team.}, in order to obtain the time--integrated energy 
spectrum. As mentioned above, the BAT detector is well calibrated only in the 
15--150 keV energy range.  Therefore, we used the 60 BAT channels in this range, 
with the default binning. 
The spectral fit to the different photon models mentioned above was performed with
the heasoft package XPSEC (version 11.3.1).

\begin{table*} 
\begin{tabular}{lccclllc} 
 
\hline

GRB      &  Lg\nor[sd]  &  $\slope$[sd]  &  Lg\ep[sd]  & $\hspace{0.09in} e_1
\hspace{0.41in} {\bf n_1}$ & $\hspace{0.09in} e_2 \hspace{0.41in} {\bf n_2}$ & $\hspace{0.09in}e_3 
\hspace{0.41in} {\bf n_3}$  & $\chi^2_{\rm red}$\\
         &  50 keV  &   &  keV  &            &         &        &   \\  
\hspace{0.08in} (1) & (2) & (3) & (4) & \hspace{0.31in} (5)  & \hspace{0.31in} (6)  & \hspace{0.31in} (7) & (8) \\
\hline
050126 & -2.38[0.21] &  -0.75[0.44] &  2.07[0.19] &  0.022 (-0.83,-0.17,-0.53) &  0.519 (  0.39,-0.85,-0.34) &  0.092 ( 0.39,0.49,-0.78) & 1.38\\
050223 & -2.47[0.22] &  -1.50[0.42] &  1.83[0.21] &  0.022 (-0.89,-0.36,-0.29) &  0.506 (-0.44, 0.83, 0.34) &  0.140 (-0.12,-0.43, 0.90) & 0.90\\
050318 & -2.02[0.14] &  -1.32[0.26] &  1.67[0.05] &  0.010 (-0.77,-0.40,-0.49) &  0.289 ( 0.46,-0.89,-0.02) &  0.051 ( 0.43, 0.24,-0.87) & 0.80\\
050401 & -1.56[0.08] &  -1.23[0.19] &  2.27[0.23] &  0.011 (-0.96,-0.11,-0.25) &  0.306 (-0.27, 0.61, 0.74) &  0.056 (-0.07,-0.78, 0.62) & 1.05\\
050505 & -2.20[0.14] &  -0.95[0.31] &  2.10[0.16] &  0.016 (-0.87,-0.16,-0.46) &  0.366 ( 0.37,-0.84,-0.41) &  0.070 ( 0.32, 0.53,-0.79) & 0.75\\
050525 & -0.62[0.03] &  -0.98[0.07] &  1.91[0.01] &  0.002 (-0.73,-0.20,-0.65) &  0.012 ( 0.59, 0.30,-0.75) &  0.077 (-0.35, 0.93, 0.10) & 0.34\\
050603 & -1.27[0.07] &  -0.97[0.17] &  2.51[0.31] &  0.010 (-0.98,-0.04,-0.21) &  0.358 (-0.20, 0.46, 0.86) &  0.049 (-0.06,-0.89, 0.46) & 0.90\\
050803 & -2.60[0.18] &  -0.99[0.37] &  1.99[0.15] &  0.018 (-0.85,-0.23,-0.48) &  0.425 ( 0.41,-0.86,-0.32) &  0.081 ( 0.34, 0.47,-0.82) & 0.85\\ 
050814 & -2.10[0.30] &  -0.58[0.56] &  1.73[0.59] &  0.017 (-0.60,-0.27,-0.75) &  0.0.75 ( 0.65, 0.38,-0.66) &  0.628 (-0.46, 0.88, 0.05) & 0.93\\ 
050908 & -2.29[0.27] &  -1.26[0.48] &  1.65[0.08] &  0.019 (-0.76,-0.41,-0.51) &  0.550 ( 0.49,-0.87,-0.02) &  0.093 ( 0.43, 0.26,-0.86) & 0.94\\ 
050922C & -1.66[0.07] &  -1.16[0.15] &  2.35[0.22] &  0.009 (-0.97,-0.09,-0.22) &  0.274 (-0.24, 0.54, 0.81) &  0.045 (-0.04,-0.84, 0.54) & 0.87\\
051109 & -1.75[0.27] &  -1.04[0.55] &  1.92[0.17] &  0.026 (-0.82,-0.26,-0.52) &  0.628 ( 0.42,-0.88,-0.22) &  0.112 ( 0.40, 0.40,-0.83) & 0.81\\
051111 & -2.02[0.06] &  -1.17[0.14] &  2.54[0.34] &  0.008 (-0.99,-0.04,-0.15) &  0.374 (-0.15, 0.35, 0.92) &  0.040 (-0.01,-0.93, 0.36) & 0.76\\
060115 & -2.51[0.16] &  -1.13[0.32] &  1.80[0.08] &  0.013 (-0.80,-0.33,-0.51) &  0.361 ( 0.45,-0.88,-0.15) &  0.063 ( 0.40, 0.35,-0.85) & 0.80\\
060206 & -1.90[0.10] &  -1.11[0.21] &  1.89[0.06] &  0.009 (-0.83,-0.28,-0.48) &  0.237 ( 0.42,-0.88,-0.22) &  0.044 ( 0.37, 0.39,-0.85) & 0.91\\ 
060210 & -2.38[0.11] &  -1.12[0.26] &  2.07[0.14] &  0.013 (-0.89,-0.19,-0.41) &  0.307 (-0.37, 0.82, 0.43) &  0.060 (-0.25,-0.53, 0.81) & 1.06\\
060223 & -2.04[0.18] &  -1.16[0.35] &  1.80[0.09] &  0.015 (-0.80,-0.33,-0.50) &  0.400 ( 0.45,-0.88,-0.15) &  0.072 ( 0.39, 0.34,-0.85) & 0.96\\
060502A & -2.05[0.08] &  -1.19[0.18] &  2.27[0.22] &  0.010 (-0.96,-0.12,-0.25) &  0.296 (-0.27, 0.60, 0.75) &  0.054 (-0.06,-0.79, 0.61) & 1.01\\ 
060510B & -2.72[0.09] &  -1.53[0.19] &  1.99[0.17] &  0.009 (-0.94,-0.27,-0.22) &  0.267 (-0.35, 0.70, 0.62) &  0.069 ( 0.01,-0.66, 0.75) & 0.62\\
060522 & -2.37[0.23] &  -0.70[0.44] &  1.85[0.08] &  0.016 (-0.71,-0.26,-0.66) &  0.071 ( 0.55, 0.39,-0.74) &  0.501 (-0.45, 0.88, 0.13) & 1.13 \\
060605 & -2.57[0.22] &  -1.00[0.44] &  2.01[0.23] &  0.022 (-0.87,-0.24,-0.43) &  0.539 (-0.41, 0.82, 0.39) &  0.103 (-0.26,-0.51, 0.82) & 0.90\\ 
060607 & -2.52[0.09] &  -1.09[0.19] &  2.15[0.15] &  0.010 (-0.93,-0.17,-0.34) &  0.251 (-0.35, 0.74, 0.57) &  0.050 (-0.15,-0.65, 0.75) & 0.89\\
060707 & -2.18[0.19] &  -0.73[0.40] &  1.84[0.06] &  0.015 (-0.68,-0.25,-0.69) &  0.062 ( 0.59, 0.37,-0.72) &  0.443 (-0.43, 0.90, 0.10) & 1.00\\
060714 & -2.52[0.12] &  -1.77[0.24] &  1.80[0.20] &  0.012 (-0.91,-0.38,-0.15) &  0.304 (-0.40, 0.74, 0.54) &  0.138 ( 0.09,-0.55, 0.83) &  1.17\\
060904B & -2.77[0.19] &  -1.07[0.37] &  1.90[0.13] &  0.017 (-0.84,-0.29,-0.47) &  0.432 (-0.44, 0.86, 0.26) &  0.081 (-0.33,-0.43, 0.84) & 0.82\\ 
060906 & -2.16[0.16] &  -1.60[0.31] &  1.65[0.09] &  0.014 (-0.83,-0.45,-0.33) &  0.346 (-0.46, 0.89,-0.05) &  0.094 (-0.31,-0.11, 0.94) & 1.09\\
060908 & -1.84[0.08] &  -0.90[0.17] &  2.15[0.10] &  0.008 (-0.89,-0.15,-0.42) &  0.208 ( 0.36,-0.81,-0.47) &  0.039 ( 0.27, 0.57,-0.78) & 0.74\\
060927 & -2.02[0.12] &  -0.93[0.24] &  1.86[0.06] &  0.009 (-0.77,-0.27,-0.58) &  0.043 ( 0.47, 0.37,-0.80) &  0.272 (-0.44, 0.89, 0.16) & 0.98 \\ 
061222B & -2.03[0.20] &  -1.30[0.37] &  1.67[0.06] &  0.015 (-0.76,-0.40,-0.51) &  0.422 (-0.47, 0.88, 0.01) &  0.075 (-0.44,-0.24, 0.86) & 1.30\\
\hline
\end{tabular} 
\caption{Spectral fit results using the CPL photon model for the class A and B GRBs. 
The best fit values of the parameters 
and their one--dimensional standard deviations (sd) are given in columns 2--4; $e_{1,2,3}$ 
and {\bf $n_{1,2,3}$} in columns 5--7 are the semi-axis lengths and principal axes of the 
error ellipsoid, respectively. The reduced $\chi^2$ is reported in the last column (dof=57).}  
\label{CPLfit}
\end{table*}

\begin{figure*}
  \vskip -0.5 cm 
  \centerline{
  \psfig{figure=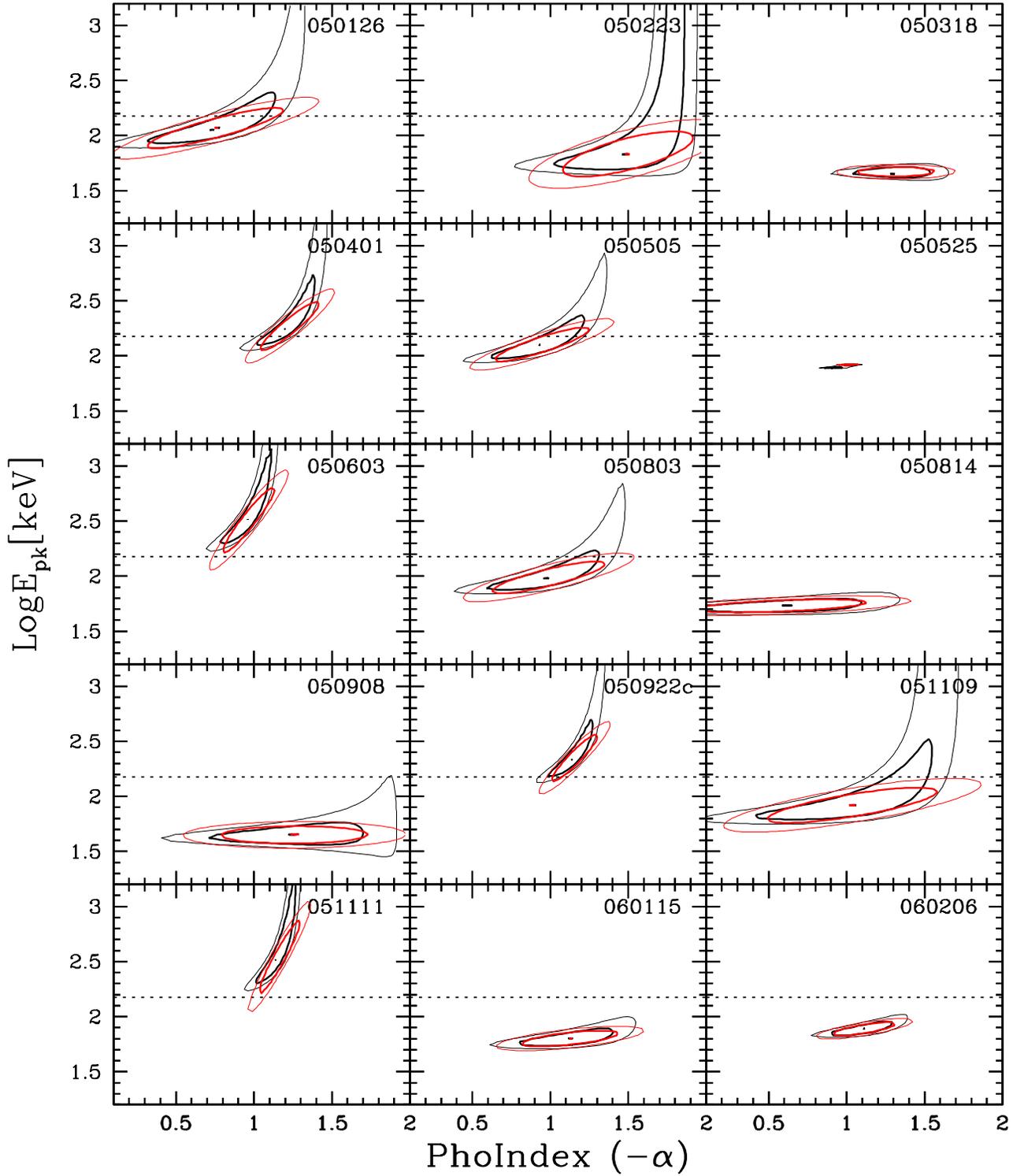,width=20.5cm,height=20cm}}
\caption{Error CL contours projected in the plane of the pair of variables 
Log\ep--(-\slope) for each one of the bursts in samples A and B (CPL model). 
The irregular contours (black lines in the electronic version) were calculated 
with the XSPEC command $steppar$, while the elliptical contours (red
lines) were constructed from the variances and principal axes given by the 
XSPEC command $fit$. Inner (thick line) and outer (thin line) contours in each 
case are for $\Delta\chi^2=1$ and 2.3, respectively (see text). The
dotted line indicates the level of 150 keV. }
\label{Epk_PI1}
\end{figure*}

\begin{figure*}
  \vskip -0.5 cm 
  \centerline{
  \psfig{figure=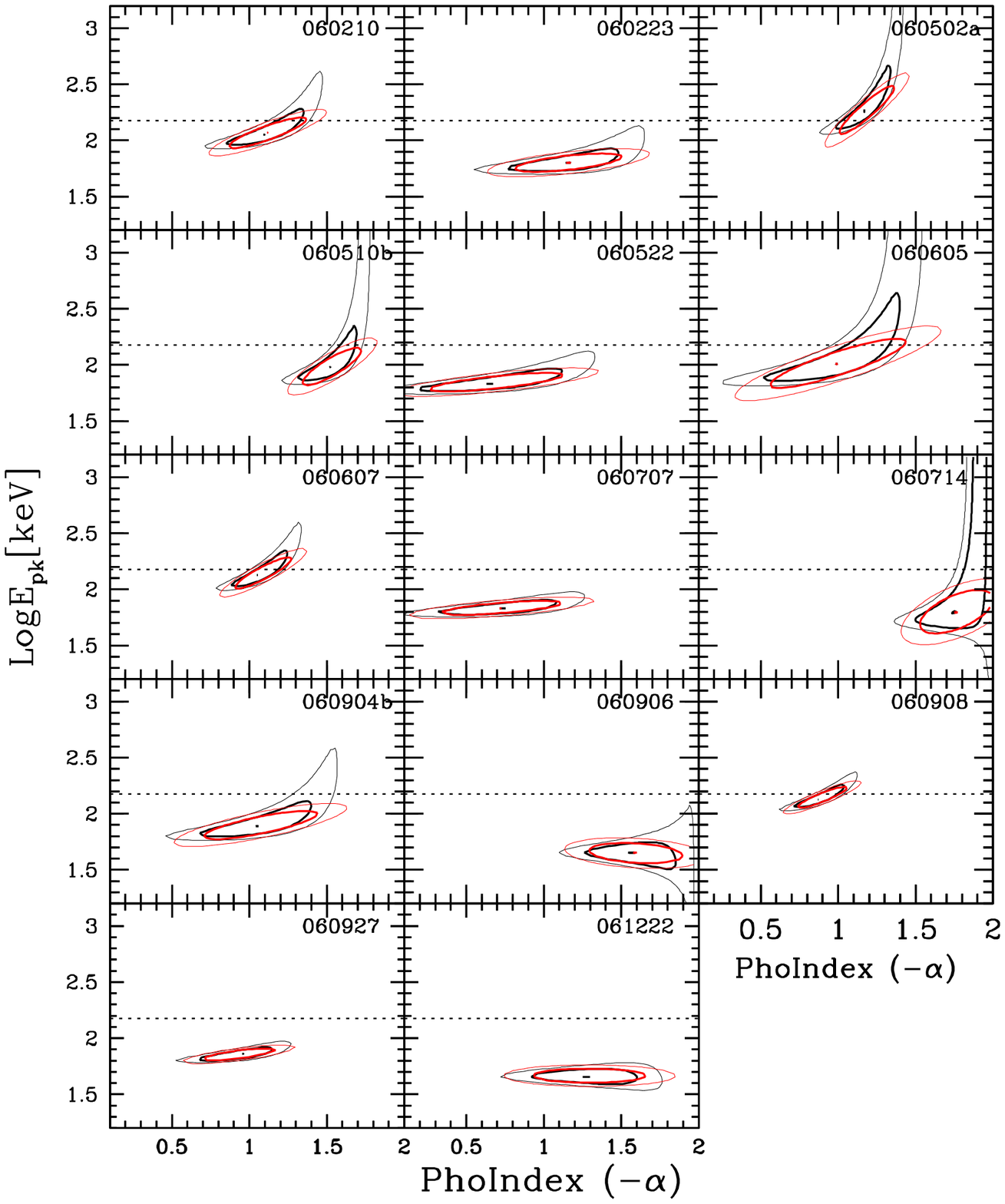,width=20.5cm,height=20cm}}
\caption{Continues Fig. \ref{Epk_PI1}}
\label{Epk_PI2}
\end{figure*}


For the GRBs for which the (CPL) \ep~ may be estimated,
we calculate the fluence, \fl, corresponding to the observed 
spectral band (15--150 keV) only with the aim of completeness.
In addition the rest isotropic--equivalent energy, \eiso, is 
calculated from the observer bolometric fluence $\fl_B$ 
corresponding to the spectral range 1--10000 keV at rest:
\begin{equation}
\eiso=\frac{ 4\pi \fl_B d_L(z)^{2} }{(1+z)},
\label{isoE}
\end{equation}
where $d_L(z)$ is the cosmology--dependent luminosity distance.

\subsection{Errors from the fit and error propagation}

A main concern of the spectral deconvolution of BAT spectra is the expected 
large uncertainty and correlations among the fitted spectral parameters. 
Consequently, a careful interpretation of the errors is crucial to make
usable and reliable the spectral analysis of \sw\ GRBs.

First, as mentioned above, we have carried out the fits using Log\nor\ 
(PL model), and Log\nor\ and Log\ep\ (CPL model), instead of the corresponding
linear values. The errors in the linear parameters
are very asymmetric and correlated among them in a very complex way; 
the corresponding regions at a given CL, as plotted with 
the XSPEC command $steppar$ in the plane of two parameters, have shapes 
strongly deviating from ellipses. 
Instead, the errors and the CL regions when the fits are carried out
with logarithmic quantities, tend to be more symmetric and
ellipse--shaped, respectively. 
This facilitates the handling of the errors to calculate composite quantities and 
correlations among the GRB parameters.

The $fit$ command of XSPEC, generates ellipsoidal confidence intervals
that approximate the complex regions of joint variation of the parameters.
These ellipsoids are characterized by the semi--axis lengths, $e_i$ 
(they are the square root of the variances), where $i$ is for each axis,  
and the corresponding principal axes (unitary vectors {\bf $n_i$}). 
We use these data to calculate the covariance matrix, $C_{ij}$, and the
elliptical boundary of the desired CL region in the subspace of interest 
\citep[see e.g.,][]{Press}. The diagonal elements of the matrix, $C_{ii}$, 
are the variances of each model 
parameter (two for the PL model and three for the CPL model), while the
off--diagonal elements (covariances) show the correlations among the parameters.
 
The error propagation from the spectral quantities (Log\nor, \slope, Log\ep) 
to the composite quantities Log\epr\ (being $\epr$ the peak energy 
at rest $\epr =\ep\times$ [1+z)]) and Log\eiso~ is performed in two
steps.
First, we have extracted sets of spectral quantities by a Monte Carlo
method according to their elliptical CL. Then, for each set we have
calculated the corresponding composite quantities as well as their
covariance matrix averaging on the overall extractions. 
From this last covariance matrix the elliptical CL of the
composite quantities is obtained.

\begin{table*} 
\begin{tabular}{lcccll}
\hline
GRB      &  \fl[s.d]  &  Lg$E_{\rm iso}^\prime$[sd] & Lg$E_{\rm
pk}^\prime$[sd] &  $\hspace{0.11in} e_1 \hspace{0.25in} {\bf n_1}$ & 
$\hspace{0.11in} e_2 \hspace{0.25in} {\bf n_2}$ \\
        & $ 10^{-7}$erg cm$^{2}$   &    $10^{50}$erg  s$^{-1}$  &
keV                &     &     \\
\hspace{0.08in} (1) & (2) & (3) & (4) & \hspace{0.25in} (5) & \hspace{0.25in} (6) \\
\hline
050126  &   8.55[1.82] & 1.70[0.18] & 2.42[0.18] & 0.062 (0.73,-0.69) & 0.247 ( 0.69, 0.73)\\
050223  &   6.14[0.83] & 1.00[0.20] & 2.04[0.21] & 0.113 (0.72,-0.69) & 0.266 ( 0.69, 0.72)\\
050318  &  12.90[0.54] & 2.07[0.11] & 2.06[0.04] & 0.107 (1.00, 0.03) & 0.044 (-0.03, 1.00)\\
050401  &  89.41[7.24] & 3.51[0.14] & 2.86[0.23] & 0.042 (0.86,-0.51) & 0.261 ( 0.51, 0.86)\\
050505  &  25.79[3.06] & 3.13[0.12] & 2.82[0.16] & 0.040 (0.79,-0.62) & 0.195 ( 0.62, 0.79)\\
050525  & 158.20[1.59] & 2.34[0.01] & 2.12[0.01] & 0.018 (0.79, 0.61) & 0.008 (-0.61, 0.79)\\
050603  &  74.60[8.18] & 3.55[0.22] & 3.10[0.30] & 0.061 (0.81,-0.59) & 0.372 ( 0.59, 0.81)\\
050803  &  20.83[2.57] & 1.13[0.14] & 2.14[0.15] & 0.044 (0.73,-0.68) & 0.199 ( 0.68, 0.73)\\
050814  &  14.60[1.16] & 2.94[0.11] & 2.53[0.06] & 0.114 (0.93, 0.36) & 0.047 (-0.36, 0.93)\\
050908  &   4.36[0.46] & 2.25[0.16] & 2.29[0.08] & 0.156 (1.00, 0.06) & 0.079 (-0.06, 1.00)\\
050922C &  16.97[0.94] & 2.63[0.14] & 2.86[0.22] & 0.034 (0.85,-0.53) & 0.257 ( 0.53, 0.85)\\
051109  &  35.27[5.85] & 2.80[0.19] & 2.44[0.16] & 0.238 (0.77, 0.64) & 0.077 (-0.64, 0.77)\\
051111  &  37.32[3.59] & 2.80[0.22] & 2.96[0.34] & 0.056 (0.84,-0.54) & 0.399 ( 0.54, 0.84)\\
060115  &  16.04[1.07] & 2.78[0.10] & 2.46[0.07] & 0.121 (0.84, 0.54) & 0.044 (-0.54, 0.84)\\
060206  &   8.29[0.35] & 2.59[0.06] & 2.59[0.06] & 0.086 (0.72, 0.70) & 0.024 (-0.70, 0.72)\\
060210  &  69.24[3.74] & 3.52[0.10] & 2.76[0.14] & 0.030 (0.81,-0.59) & 0.170 ( 0.59, 0.81)\\
060223  &   6.54[0.52] & 2.56[0.12] & 2.53[0.08] & 0.139 (0.86, 0.50) & 0.052 (-0.50, 0.86)\\
060502A &  22.87[1.66] & 2.42[0.14] & 2.67[0.22] & 0.039 (0.85,-0.53) & 0.259 ( 0.53, 0.85)\\
060510B &  38.64[2.88] & 3.53[0.13] & 2.76[0.17] & 0.046 (0.80,-0.59) & 0.207 ( 0.59, 0.80)\\
060522  &  10.49[1.04] & 2.78[0.09] & 2.63[0.08] & 0.120 (0.74, 0.67) & 0.036 (-0.67, 0.74)\\
060605  &   5.33[1.12] & 2.31[0.20] & 2.69[0.22] & 0.074 (0.74,-0.67) & 0.289 ( 0.67, 0.74)\\
060607  &  24.91[1.58] & 2.94[0.10] & 2.76[0.15] & 0.026 (0.83,-0.55) & 0.173 ( 0.55, 0.83)\\
060707  &  16.26[1.13] & 2.70[0.08] & 2.48[0.06] & 0.097 (0.81, 0.59) & 0.036 (-0.59, 0.81)\\
060714  &  30.58[3.96] & 3.11[0.15] & 2.37[0.20] & 0.111 (0.85,-0.53) & 0.219 ( 0.53, 0.85)\\
060904B &  14.78[1.91] & 1.44[0.13] & 2.13[0.13] & 0.174 (0.71, 0.71) & 0.047 (-0.71, 0.71)\\
060906  &  23.82[1.92] & 3.16[0.14] & 2.32[0.09] & 0.142 (0.98,-0.20) & 0.087 ( 0.20, 0.98)\\
060908  &  26.56[1.11] & 2.78[0.07] & 2.68[0.10] & 0.016 (0.81,-0.59) & 0.123 ( 0.59, 0.81)\\
060927  &  11.66[0.46] & 2.92[0.06] & 2.68[0.06] & 0.077 (0.72, 0.70) & 0.022 (-0.70, 0.72)\\
061222B &  22.35[1.23] & 2.97[0.14] & 2.31[0.06] & 0.137 (1.00, 0.01) & 0.064 (-0.01, 1.00)\\
\hline
\end{tabular}
\caption{Calculated fluence and rest--frame isotropic and peak energies for our 
GRB groups A and B. The corresponding one--dimensional standard deviations are given 
between square brackets. The last 2 columns give the estimated semi--axis lengths ($e_{1,2}$)
and principal axes ({\bf $n_{1,2}$}) of the error ellipse associated to the pair of quantities 
Log\epr\ and Log\eiso. Note that the errors in  most of the cases are correlated significantly 
(see also the plotted ellipses in Figs. 4 and 5).}  
\label{composite}
\end{table*}

\section{Results}

In Table \ref{PLfit} we present the results from the spectral fit to the PL model
for the whole sample of 47 long GRBs. Columns 1 to 4 give the burst name, the
class group of the burst, $z$, and \tdur. 
Columns 5 and 6 give the best fit values of the normalization, Log\nor, and the 
photon index, \slope, with the corresponding one--dimensional symmetric standard
deviations in brackets, as obtained with the XSPEC $error$ command. In columns 7 
and 8 we report the square root of the variances, $e_i$, together with the unitary vectors
$\bf n_i$ (principal axes), which provide respectively the semi--axis lenghts and 
directions of each axis ($i=1,2$) of the joint error ellipse. The reduced $\chi^2$ 
is reported in column 9. 

As it may be appreciated from Table \ref{PLfit}, the best--fit \slope\ is smaller 
than $-2$ for 7 bursts out of the 47 (GRB 060906 is marginal and it will be 
recovered by the CPL fitting). 
Most likely, the \sp\ spectra of these GRBs peak at energies below 15 keV.
They could  be related to the X--ray rich events and X--ray Flashes 
discovered by $Beppo$SAX and \hete~ (Kippen et al. 2001; Sakamoto et al. 2005).

We have also fitted the 47 GRB spectra to the CPL model and found
that for 18 bursts the fit produces too large uncertainties
and/or unphysical solutions. For these 18 cases the PL model is 
definitively more reliable than the CPL one.
According to the results from our spectral fits (judged by the joint
68\% CL contour errors), we classify the 47 \sw\
GRBs analysed here in four classes:

{\it Class A --} The CPL--model fit is acceptable and \ep~ is within the 15--150 
keV range. 

{\it Class B --} The CPL--model fit is acceptable, but $\ep > 150$ keV.

{\it Class C --} the CPL--model fit gives unreliable results, while
the PL--model fit is acceptable and gives \slope$> -2$, suggesting
that \ep\ is at energies $ > 150$ keV.

{\it Class D --} The CPL--model fit gives unreliable results, while the PL--model 
fit is acceptable and gives \slope$< -2$, suggesting that \ep\ is smaller than 15 keV.

The class of each GRB is indicated in the second column of Table \ref{PLfit}.  
This classification can be easily understood after a visual inspection
of the plots shown in Appendix and figures \ref{Epk_PI1} and \ref{Epk_PI2}.   
In Figs. \ref{spectra1}, \ref{spectra2}, and  \ref{spectra3} of the
Appendix we present the deconvolved time--integrated \sp\ [or $E^2f(E)$] 
observed spectra of the 47 \sw\ GRBs analysed here. The CPL model was
used for the spectral deconvolution.  The error bars show the observations 
while the continuous lines show the fitted curves.  
The spectra of all the events that we classify in groups A and B show
some curvature, evidence of a peak in the $E^2f(E)$ distribution. Nevertheless,
in most of the cases the formal $F-$test analysis says that the spectral
fits from the PL model to the CPL one (which introduces one more
parameter, \ep) do not improve significantly. Therefore, one expects
large uncertainties in the determination of \ep~ for most of the events
analyzed here. {\it This is why it is important to carry out a
carefully statistical analysis of the errors and to adequately 
propagate them for calculating composite quantities and inferring 
correlations.}

Figures \ref{Epk_PI1} and \ref{Epk_PI2} show different contours of the CLs projected 
in the plane of the pair of variables $Log\ep$ vs. ($-\slope$) for each 
burst in groups A and B (CPL model). The irregular CL contours (black lines
in the electronic version) were calculated with the XSPEC command $steppar$, 
while the elliptical CLs (red lines) were constructed from the joint variances and 
principal axes given by the XSPEC command $fit$ (see \S\S 2.3; columns 5, 6, and
7 in Table \ref{CPLfit} below).  Thick lines show the CLs for  $\Delta\chi^2=1$, 
while thin lines show the CLs for $\Delta\chi^2=2.3$.  
For $\Delta\chi^2=1$, the projections of the CL regions in each axis
enclose $\approx 68\%$ of the one--dimensional probability for the given 
parameter (one standard deviation), while for $\Delta\chi^2=2.3$, the plotted 
region encompasses $\approx 68\%$ of the probability for the {\it joint variation} 
of the two parameters (one standard deviation) \citep[e.g.,][]{Press}.
As can be appreciated, the ellipses are a reasonable description of the CL regions 
in most cases, with the great advantage that they are handled analytically.

We see from Figs. \ref{Epk_PI1} and \ref{Epk_PI2} that for many of the GRBs, the  
Log\ep\ and \slope\ spectral parameters are correlated (inclined, thin ellipses). 
A similar situation arises for the other combination of parameters. Therefore, for 
this kind of data, a simple statistical quantification of the errors, such as the 
one--dimensional standard deviation, is not enough.  This is why we have worked out 
the full covariance matrix calculated from the error ellipse.
We use it to propagate errors in the calculation of \epr\ and \eiso\ (\S\S 2.3). 
{\it This is a crucial step when using the data, for instance, for 
establishing correlations among the spectral or composite quantities} (\S\S 3.3).    

The CPL--model fittings for GRB groups A and B are presented in Table \ref{CPLfit}. 
Column 1 gives the burst names. Columns 2, 3, and 4 report the best fit values of 
Log\nor, \slope, and Log\ep~ with their corresponding one--dimensional standard 
deviations, as calculated with the XSPEC $error$ command. In columns 5, 6, and 7 we 
report the lengths $e_i$ of each semi--axis together with the unitary vectors
$\bf n_i$ (principal axes), which provide the directions of each axis
($i=1,2,3$). The reduced $\chi^2$ is reported in column 8 (dof = 57).

Finally, in Table \ref{composite} we report the calculated values for \fl, 
Log\eiso, and  Log\epr\ , as well as the corresponding 
propagated errors for the samples A and B, using the CPL--model fit results
(see \S 2.3 for the used error propagation procedure).
Column 1 gives the burst name. Columns 2 and 3 give respectively the values of \fl~ 
(in the observed range 15--150 keV) and of \eiso\ (extrapolated to the rest--frame
energy range 1--10000 keV) corresponding to the best fit spectral parameters. 
Column 4 gives the value of the rest--frame peak energy, \epr. The one--dimensional
standard deviationss of \fl, \eiso, and \epr\ are given within square brackets.  
However, due to the high correlation among the errors the simple standard deviation 
is not enough to characterize statistically the errors. Thus, in columns 5 and 6 
we report the error ellipse elements (semi--axis lenghts, $e_i$, and principal
axes $\bf n_i$, $i=1,2$) associated to the pair of quantities Log\epr\ and Log\eiso. 
It is important to remark that 
the correlation between the uncertainties of Log\epr\ and Log\eiso\ is not entirely 
due to the correlation among Log\nor, \slope, and Log\ep. For instance an increase 
in Log\ep\ alone is able to produce an increase in both Log\epr\ and Log\eiso. 
This effect will turn out to be helpful in the next section, where we will
study how Log\epr\ and Log\eiso\ do correlate in the entire GRB sample.

The median and dispersion (quartiles) of the best--fit CPL parameter \slope~ 
for the 24 + 5 classes A+B GRBs are $-1.12^{+0.15}_{-0.14}$, which is in
agreement with the results from BATSE bright--GRB \citep[][]{Kaneko06} and 
\hete~ GRB \citep{Barraud03,Sakamoto05} time--integrated spectra. Our \slope\ 
distribution is in particular similar to the one from \hete, with only $\sim 25\%$
of the bursts having $\slope\la -1.2$.
The median and quartiles of the Log\ep\ [keV] parameter
are $1.91^{+0.19}_{-0.11}$. Unfortunately, due to the
low energy limit of the \sw--BAT, the sample analysed here covers only
the $\sim 15\%$ fraction of bursts in the (CPL--model) \ep\ distribution inferred
for BATSE bright bursts  \citep[][]{Kaneko06}. 

The distribution of \epr\ (at rest) is much broader than the one of \ep\
(the median and quartiles of Log\epr[keV] are $2.54^{+0.22}_{-0.21}$). 
This is expected due to the broad distribution in redshfits of the \sw\ bursts
analysed here.  For our 24 + 5 classes A+B GRBs, the $z$ distribution has a
broad maximum at $z\sim 3-4$; the median and quartiles are $3.08^{+0.82}_{-1.5}$. 
The  median and quartiles of Log\eiso\ [$10^{50}$ erg] for our 24 + 5 classes A+B 
GRBs are $2.78^{+0.30}_{-0.44}$.  For the 8 long--duration \sw\ 
GRBs in common with the compilation presented in Amati (2006b), the values
of \epr~ reported here and in that paper are similar, but those of \eiso~ are 
systematically larger in Amati (2006b). The latter difference is probably due to 
the fact that Amati (2006b) did not use for most of those GRBs the spectral 
information from the \sw~ satellite but from other space missions (Konus, \hete),
which allowed to find a fitting with the Band model; as will be discussed 
in \S\S 3.3, this implies typically a larger \eiso~ with respect to the case a 
CPL model is used.

\begin{figure}
\vskip -0.7 true cm
\centerline{ \psfig{file=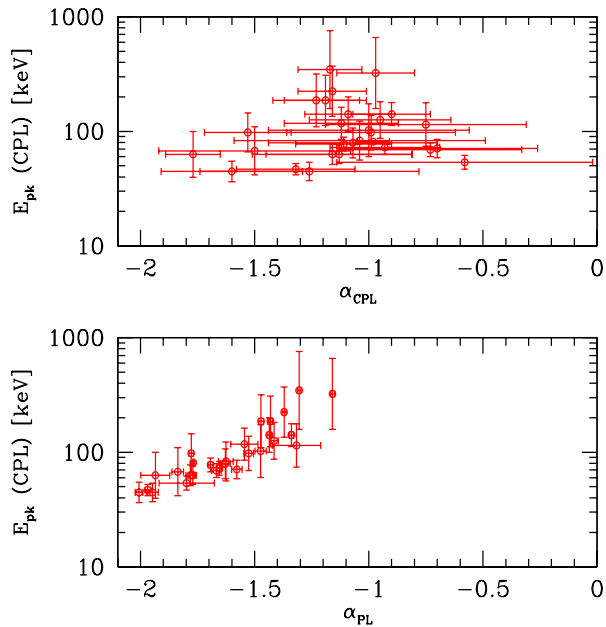,width=9.5cm,height=10cm} }
\vskip -0.9 true cm
\caption{Top panel: the observed \ep\ as derived by the CPL fit as a
function of the spectral index $\alpha_{\rm CPL}$ of the same fit.
Bottom panel: \ep\ (as derived by the CPL fit) as a
function of the spectral index $\alpha_{\rm PL}$ of the single power law fit.
The correlation of \ep\ with $\alpha_{\rm PL}$ is induced by \ep\ 
entering in the \sw\ energy band (see text), and $cannot$ be used
to infer \ep\ when knowing only $\alpha_{\rm PL}$. 
}
\label{epalpha}
\end{figure}

\subsection{Bursts detected also by other instruments}

Among the bursts listed in Tables 2 and 3 there are four GRBs
which have been detected by other instruments besides \sw.
Three of them have been detected by Konus--Wind (GRB 050401, 050603 and 051109A)
and one by  \hete~ (GRB 050922C).

For GRB 050401 the Konus--Wind results reported by Golenetskii et al. (2005a)
concern the spectral parameters for the two peaks displayed by its light curve,
fitted with a Band model in the energy range [20--2000] keV. The first peak had 
$\alpha=-1.15\pm 0.16$, $\beta= -2.65\pm 0.31$ and $E_{\rm pk}=132\pm 16$ keV, 
while the second peak is described by
$\alpha = -0.83 \pm 0.21$, $\beta = -2.37 \pm 0.14$ and $E_{\rm pk} = 119 \pm 26$ keV.
Our analysis with the  \sw~ data (see Tab. 3) for the time integrated spectrum
and a CPL model yielded $\alpha=-1.23\pm 0.19$ and $E_{\rm pk}=186^{+130}_{-76}$ keV.
Within the (relatively large) errors, the results are consistent.

The Konus--Wind data of GRB 050603 have been fitted by Golenetskii et al. (2005b)
with a Band model in the [20--3000] keV energy range, with 
$\alpha = -0.79 \pm 0.06$, $\beta = -2.15 \pm 0.09$ and $E_{\rm pk} =349\pm 28$ keV.
Our analysis with a CPL model gives 
$\alpha=-1.23\pm 0.19$ and $E_{\rm pk}=323_{-164}^{+340}$ keV.
The derived low energy spectral index is somewhat softer, while the
value of $E_{\rm pk}$ is consistent.

GRB 050922C was detected by \hete~ with the FREGATE instruments in the 7--30, 7--80, 
and 30--400 keV bands. It was fitted with a CPL law model by Crew et al. (2005) with 
$\alpha=-0.83_{-0.26}^{+0.23}$ and $E_{\rm pk}=130.5_{-26.8}^{+50.9}$ keV, to be compared 
with our values of $\alpha=-1.16\pm 0.15$ and $E_{\rm pk}=224_{-89}^{+148}$ keV.
Again, within the errors, there is consistency.

GRB 051109, detected by Konus--Wind and fitted with a CPL model by
Golenetskii et al. (2005c) in the  [20--500] keV energy range, showed 
$\alpha=-1.25_{-0.59}^{+0.44}$ and $E_{\rm pk} = 161_{-58}^{+224}$ keV.
This should be compared with our $\alpha=-1.04\pm 0.55$ and
$E_{\rm pk}=83_{-27}^{+40}$ keV.
Even in this case there is consistency, within the relatively large errors.

\subsection{Peak energy vs spectral index}

In Fig. \ref{epalpha} we show \ep, derived with the CPL fit, as a function of
the spectral index $\alpha_{\rm CPL}$ of the CPL fit (top panel) and as 
a function of $\alpha_{\rm PL}$, the index derived when fitting with a single 
power law. As can be seen, although there is no relation between \ep\ and 
$\alpha_{\rm CPL}$, a correlation appears when fitting with a single power law.
This correlation, however, has no physical meaning, since it is the result of 
the attempt of the single power law to account for the data point above the peak,
with smaller flux (in \sp). As \ep\ decreases, a larger fraction of data
lies above the peak, inducing the single power law to steepen. This also means 
that this effect vanishes when \ep\ is outside the \sw\ energy range, therefore
{\it if we do not know \ep, we cannot use this
correlation to infer it}.
Furthermore, note the difference in the derived values of
the spectral indices.
As expected for these bursts for which we could derive \ep, 
the index $\alpha_{\rm PL}$  is systematically softer than
$\alpha_{\rm CPL}$.
Given the obvious result of this subsection, we have considered
unnecessary to show the upper panel of Fig. \ref{epalpha} with the
$68\%$ CL error ellipses. A complete treatment of the errors in this
sense confirms the previous result.

\begin{figure*}
\centerline{
 \psfig{file=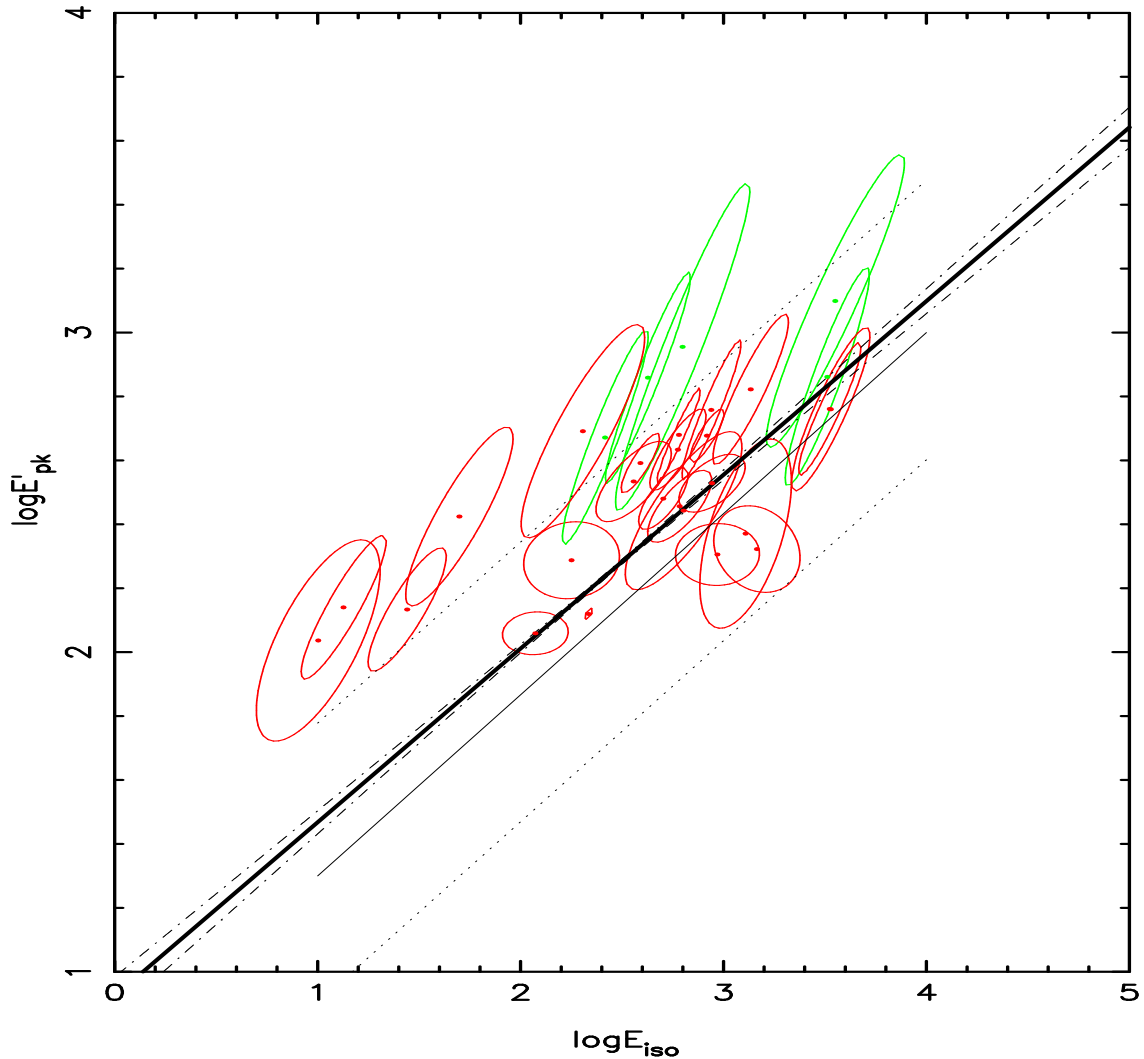,width=15cm,height=14cm}}
 \caption{Correlation between \epr~ and \eiso~ (rest frame) as obtained from 
spectral fits using the CPL model for the class A (red ellipses) 
and B (green ellipses) bursts. The ellipses correspond to the joint $68\%$ CL 
error regions of each data point. The dotted lines encompass the $3\sigma$ scatter 
of the updated Amati correlation presented in \citet{Ghirla07} for 49 GRBs; the thin 
solid line is the best fit for that sample. The thick solid line is the best fit 
that we find for the \sw~ data presented here taking into account the correlated 
errors, and the dot--dashed line is for the 1$\sigma$ uncertainty, computed in the 
barycentre of the data points.}
 \label{amati}
\end{figure*}

\subsection{Correlation between \epr\ and \eiso}

In Fig. \ref{amati}, \epr\ is plotted vs \eiso\ for the class A (red--line ellipses) 
and class B (green--line ellipses) \sw~ GRBs. The ellipses correspond to the joint 
$68\%$ CL error 
regions calculated as explained in \S\S 2.3. The dotted straight lines encompass the 
$3\sigma$ scatter of the updated Amati correlation presented in \citet{Ghirla07}
for 49 GRBs; the central thin solid straight line is the best fit for that sample.
A visual inspection shows that the \sw~ data analyzed in this
  paper are indeed correlated in the \epr--\eiso~ diagram.
  Taking into account the uncertainties, no GRB may be classified as
  outlier.
Eight GRBs show their central point above the $3\sigma$ strip, but taking
into account the $68\%$ CL contours as well as the fact that the spectral 
fitting was performed using a CPL model instead of the Band one (see below),
 make such GRBs compatible with the $3\sigma$ strip.
Notice that the 5 (marginal) class B events are not outliers. Therefore,
we may include them for a fit along with the 24 class A events.

The thick solid line in Fig. \ref{amati} is the best fit that we find
for the 29 class A and B GRBs. 
The linear fit has been performed by an iterative minimum $\chi^2$ method, 
where the coordinate origin of the data is shifted to the barycentre,
and the uncertainty component of each datapoint is calculated from the 
corresponding error ellipse in the orthogonal direction of the straight 
fitted line.
The dot--dashed line curves describe the standard error of the mean 
values of the fit parameters computed in the barycentre of the data points.
The best fit in logarithmic quantities is:
\begin{equation}
{\rm Log}\left(\frac{\epr}{\rm keV}\right) = (2.25 \pm 0.01) + (0.54 \pm 0.02) 
{\rm Log}\left( \frac{\eiso}{10^{52.44}{\rm erg}}\right) 
\label{corr} 
\end{equation}
We point out the weight of GRB050525 in determining the best fit
due to its very--reduced spectral parameter uncertainties. We have also carried out
the linear fit for only the 24 class A GRBs:
\begin{equation}
{\rm Log}\left(\frac{\epr}{\rm keV}\right) = (2.23 \pm 0.01) + (0.54 \pm 0.02) 
{\rm Log}\left( \frac{\eiso}{10^{52.42}{\rm erg}}\right) 
\label{corr1} 
\end{equation}
As expected, the fit is very similar to the one that includes the 5 class B GRBs.

Thus, a correlation between \epr\ and \eiso\ is confirmed for 29 long GRBs 
observed by the \sw\ satellite and homogeneously analyzed by us. The values 
of the power--law index ($\alpha\sim 0.54$) and normalization 
($K\sim 100$ keV for \eiso~ given in units of $10^{52}$erg, $\epr=K\eiso^\alpha$) 
corresponding to the best fit, lie in the ranges of values found with different 
previous samples (see for a review of previous studies in Amati 2006a).
It is encouraging that with the \sw\ data, as analysed here, one obtains an 
\epr--\eiso\ correlation fully consistent with the one obtained for GRBs observed 
with other satellite detectors \citep[the so--called Amati relation;][see also 
Amati 2006a and Ghirlanda 2007 for the most recent updates]{Amati02}, which had
a much broader energy (spectral) range than BAT. The errors
in the fit parameters reported in eqs. (\ref{corr}) and (\ref{corr1}) are
at the 68\% CL. These errors are somewhat larger than those reported
in Ghirlanda (2007) and Amati (2006a; note that therein the errors
are at the 90\% CL). 

In general, the scatter around the correlation obtained here is within the range
  of different previous determinations of the Amati correlation.
  The value of the reduced $\chi^2$ for the fit with the $29$ class A+B
  GRBs is $\chi^2_\nu=379/27=14$ close to the values
  $\chi^2_\nu=493/41=12$ and $\chi^2_\nu=530/47=11.3$ obtained for larger 
  samples ($43$ and 49 GRBs) by Ghirlanda et al. (2006) and Ghirlanda (2007), 
  respectively. 
  We have also estimated the variance of our fit, $s^2$ \citep[see e.g.,][Chapter 11]{Bevington}, 
  by taking into account the joint variance of the data, i.e. the elliptical CL contours. For the
fit using the 24+5 GRBs, we find $s=0.13$\ (dof=27). Note that for calculating
the deviations, we use orthogonal distances to the best--fitting line rather
than distances projected in the Log\epr~ (Y) axis. An estimate of the vertical
1$\sigma$ spread of the data around the best--fitting model would be
$\sigma_{\rm Log\epr}= 0.15$, in agreement with previous estimates
\citep[see e.g.,][]{amati06a}.  

The virtue of the \sw\ sample analysed here is its homogeneity, 
in the sense of observations, reduction, and analysis. Previous
samples used to infer the \epr--\eiso~ correlation were 
constructed from different detectors, with an information collected from 
different authors, who used different data processing and assumptions. 
This heterogeneity in the data should introduce a spurious scatter.
In spite that the data used here have large observational errors, the
dispersion of our fit lies within the range of fits reported previously. 
This is namely due to the homogeneity of the sample and because, having taken
into account the whole error ellipses associated to the data points, 
it results that the major--axis orientation of most of these 
ellipses is close to the direction of the correlation line. 

Examining in more detail Fig. \ref{amati}, one sees that the normalization of the 
correlation found here [eq. (\ref{corr})] is slightly higher (by $\approx 0.1$ dex
at the sample median value of \eiso) than the previously established one. This is 
indeed expected. Due to the narrow energy range of BAT, we had to force 
the GRB spectra to be fitted by the CPL model, while most spectra could have been
better described by the Band model, if higher energies had been detected. It is  
known that when fitting the CPL model to a spectrum whose shape is actually described 
by the Band model, then (i) \ep\ is larger than it would otherwise be, and (ii) 
the low--energy power--law index $\alpha$ is more negative than it would otherwise 
be \citep{Band93, Kaneko06}. On the other hand, (iii) in the CPL model the \sp\ 
spectrum is cut--off 
exponentially after the break energy \ebr, while in the Band model the spectrum
decreases as a power law. Item (i) implies directly that \ep\ is likely 
overestimated when the CPL model is used in the spectral analysis, and items (ii) 
and (iii) imply that \eiso, which is calculated in the rest--frame energy range of 
1--10000 keV by extrapolating the fitted spectrum, is likely underestimated.

\begin{figure*}
 \centerline{
\psfig{figure=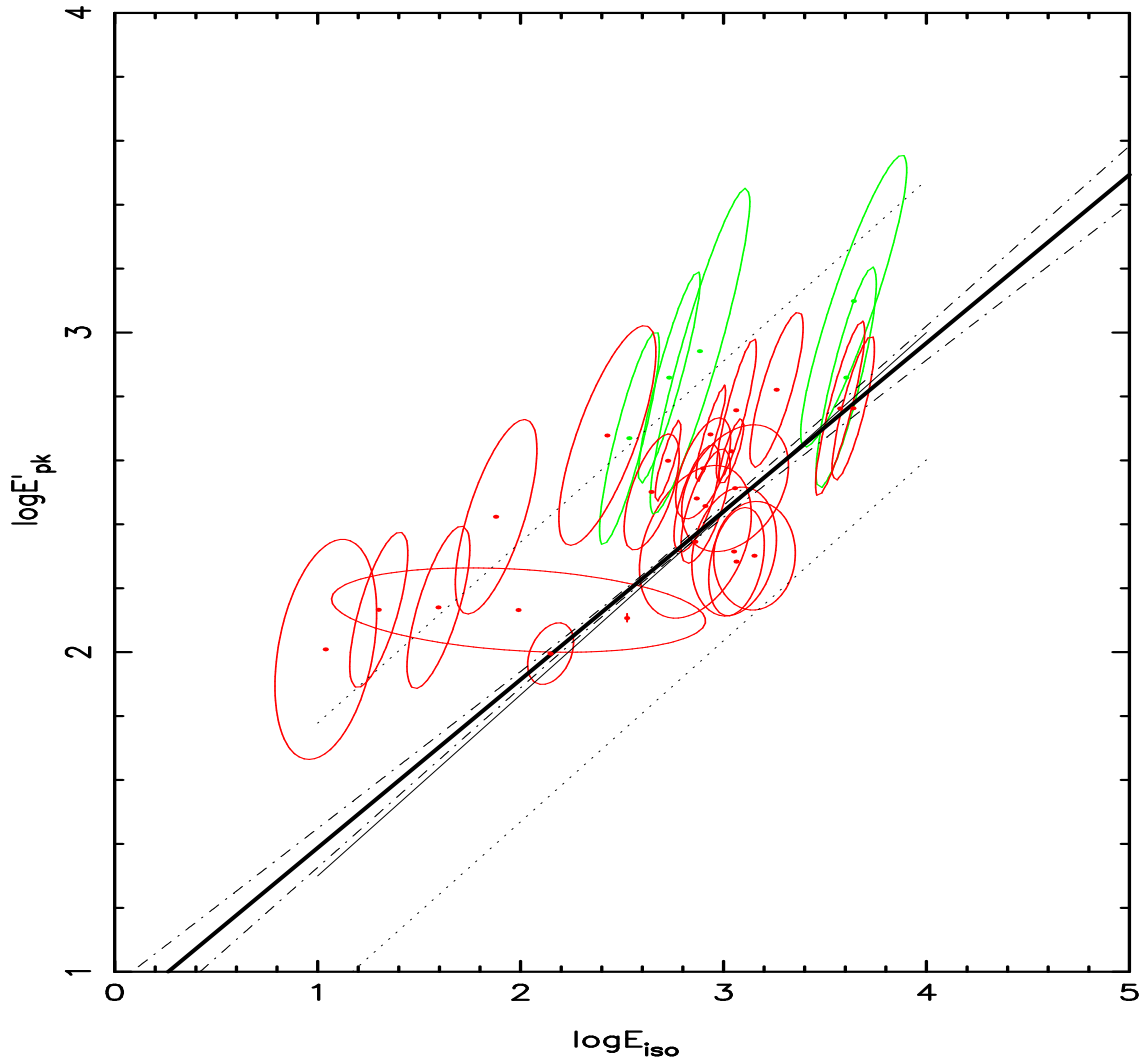,width=15cm,height=14cm}}
\caption{Same as Fig. \ref{amati} but for \epr~ and \eiso~ inferred from spectral
fits using the Band model with $\beta$ frozen to $-2.3$. Note that using the Band
model the correlation shifts to the higher (lower) \eiso~ (\epr) side with respect
to the case when the CPL model is used.}
\label{amati2}
\end{figure*}

In order to explore in a statistical sense the effect in the \epr--\eiso\ 
correlation of using the CPL model instead of the Band one for the
spectral fitting, we carried out the following experiment. 
The time--averaged spectra of the 29 long GRBs that had a reasonable CPL--model 
fit (groups A and B) were fitted with XSPEC to a Band model, but with the high--energy 
power law index $\beta$ frozen to the typical value of $-2.3$. As in the case of CPL, 
we have modified the Band model in order to perform the fit for the logarithms of the 
normalization and \ep. Then, we followed the same procedures described
in \S 2 for handling the errors, taking into account the correlation
among them. Fig. \ref{amati2} shows the final data points and their 68\% CL 
ellipses in the \ep--\eiso\ diagram. As expected, the normalization of the 
\epr--\eiso\ correlation decreases (with respect to Fig. \ref{amati}), now being
almost equal to the one of the updated Amati relation given in \citet{Ghirla07}
for the range of \eiso\ values studied here. The best fit in logarithmic quantities is:
\begin{equation}
{\rm Log}\left(\frac{\epr}{\rm keV}\right) = (2.23 \pm 0.01) + (0.53 \pm 0.04)
{\rm Log}\left(\frac{\eiso}{10^{52.61}{\rm erg}}\right)
\label{corr_band} 
\end{equation}

\section{Summary and Conclusions}

We have analysed homogeneously the time--integrated spectra of a relatively large
sample of long--duration GRBs with known redshift obtained by a same detector, 
the \sw--BAT instrument. We have carried out spectral fits to 47 GRBs by using 
the two--parameter PL and three--parameter CPL photon models. Due to the narrow 
spectral range of BAT, 15--150 keV, only two-- or three--parameter models can 
give reliable fits. However, even in these cases, large uncertainties are expected 
in the fitted parameters, making the study of the (highly correlated) errors 
important. We have performed Monte Carlo simulations in order to propagate the 
errors and calculate composite quantities such as \epr\ and \eiso. 
The main results and conclusions are as follow:

$\bullet$ For 29 GRBs (classes A+B) out of the 47 bursts analysed here, the CPL model, 
which identifies a peak energy \ep~ in the \sp spectrum, gives reliable fits, though 
in 5 cases (class B), the best fit value for \ep~ lies beyond the upper limit (150 keV) 
of the BAT energy band. The uncertainties in the fitted parameters are in most cases 
highly correlated among them. 

$\bullet$ The spectra of the remaining 18 bursts (classes C+D) are well fitted by the 
simple PL 
model. For 11 bursts (class C) of the 47 GRBs, the best PL--fit value of the photon index 
\slope~ is greater than $-2$, suggesting that $\ep > 150$ keV. 
The other 7 bursts (class D) show a  photon index \slope~  lower than $-2$ 
which implies a decreasing power--law \sp spectrum in the 
analysed energy range. In these cases, the peak energy could lie in the X--ray range. 

$\bullet$ The fluence, \fl, and rest--frame isotropic equivalent energy, \eiso, as 
well as their uncertainties, were calculated for the 29 (classes A+B) GRBs with
reliable fits to the CPL model. This is currently the largest {\it homogeneous} 
sample of long GRBs with determined spectral parameters and \eiso. The mean
values of the rest peak energy\epr\ and \eiso\ in the sample are $\sim
350$ keV and $\sim 4\times 10^{52}$ erg, respectively.

$\bullet$ The \ep\ as inferred from the CPL fit and 
$\alpha_{\rm PL}$ as inferred from the PL fit do correlate. 
However, this correlation is not physical, but is the result of the 
attempt of the single PL to account for the data points above the peak,
with smaller flux. As \ep\ decreases, a larger fraction of data
lies above the peak, inducing the single power law to steepen.
Therefore, the correlation {\it should not} be used to infer \ep\ when 
knowing only $\alpha_{\rm PL}$. Indeed, we show that \ep~ and  
$\alpha_{\rm CPL}$ do not correlate. 

$\bullet$ A correlation between \epr\ and \eiso\ (the 'Amati' relation) is 
confirmed for our sample [Eq. (\ref{corr})]. This correlation and
its scatter are consistent with 
the ones established previously for non--\sw\ bursts, showing that the
\epr--\eiso\ correlation is hardly an artifact of selection effects. 
The zero--point of our correlation (in Log--Log) is larger by $\sim 0.1$ dex at
the sample median value of \eiso\ than the latest updated 
``Amati''  correlation  \citep{Ghirla07}. 
This difference is expected due to the use of a CPL model for
describing the observed spectra instead of the Band model.

$\bullet$ When the Band model with the high--energy photon index frozen to 
$\beta=-2.3$ is fitted to the spectra of our 29 (classes A+B) GRBs, the zero--point of the 
resultant \epr--\eiso\ correlation becomes smaller than in the case of the CPL model. 
For the Band model, the obtained \epr--\eiso\ correlation 
is given by Eq. (\ref{corr_band}), 
which is virtually indistinct from the 'Amati'  correlation established 
previously for uneven observable data-sets from different satellites.

 Although in this work we were able to process homogeneously the spectra of a 
significant fraction of the \sw~ GRBs with known redshift, the sample is still
limited for conclusive results regarding the physical meaning of the correlations 
among the burst observable properties. During the refereeing process of this work, 
appeared a paper by \citet{Li07b}, where the author finds evidence of significant 
change in the 'Amati' correlation parameters with redshift for the heterogeneous 
sample of 48 GRBs compiled by \citet{amati06a}. The number of GRBs with known 
redshifts and full spectral information obtained homogeneously should increase in 
order to attain more conclusive results.  

\section*{Acknowledgments}

JIC and VA-R gratefully acknowledge the hospitality extended by 
Osservatorio Astronomico di Brera in Merate during research stays. 
JIC is supported by a CONACyT (Mexico) fellowship. 
We are grateful to the anonymous Referee for his/her comments,
which helped to correct some bugs in the processed data and
to improve the manuscript. This work 
was supported by the PAPIIT--UNAM grant IN107706-3 to VA-R, and by an 
italian 2005 PRIN--INAF grant. The authors acknowledge the use of the
\sw\ publicly available data as well as the public XSPEC
software package. We thank J. Benda for grammar corrections to 
the manuscript.

{}

\appendix

\section{The observed spectra}

\begin{figure*}
  \centerline{
  \psfig{figure=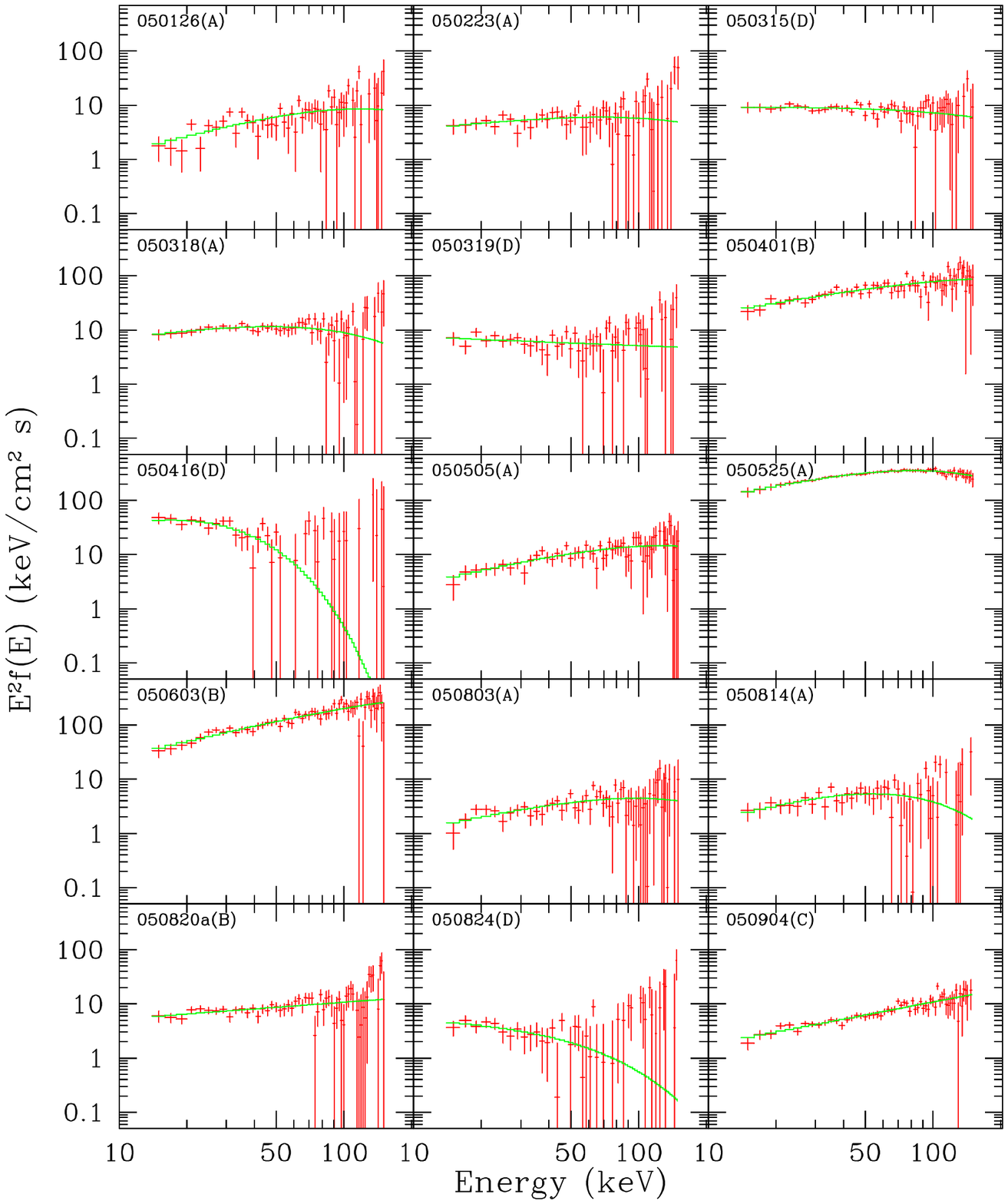,width=17.8cm,height=20cm}}
\caption{Time--averaged \sp spectra (red error bars) of the \sw~ GRBs with known 
redshift included in our sample. Continuous (green) line: best fit curve with the 
CPL photon model. The spectral flux of each burst was averaged over its integration 
time, \tdur.  }
\label{spectra1}
\end{figure*}

\begin{figure*}
  \centerline{
  \psfig{figure=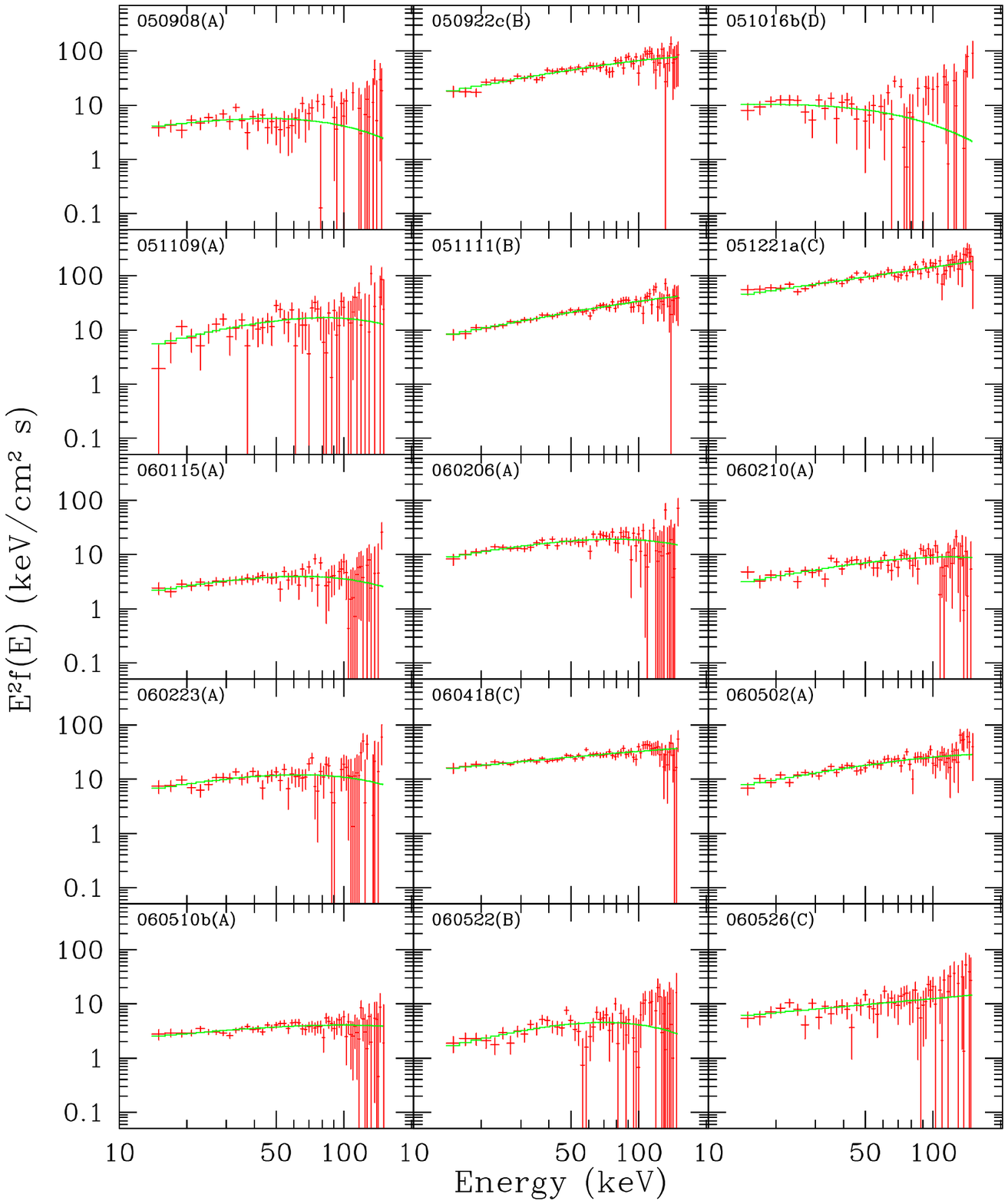,width=17.8cm,height=20cm}}
\caption{Continues Fig. \ref{spectra1}.}
\label{spectra2}
\end{figure*}

\begin{figure*}
  \centerline{
  \psfig{figure=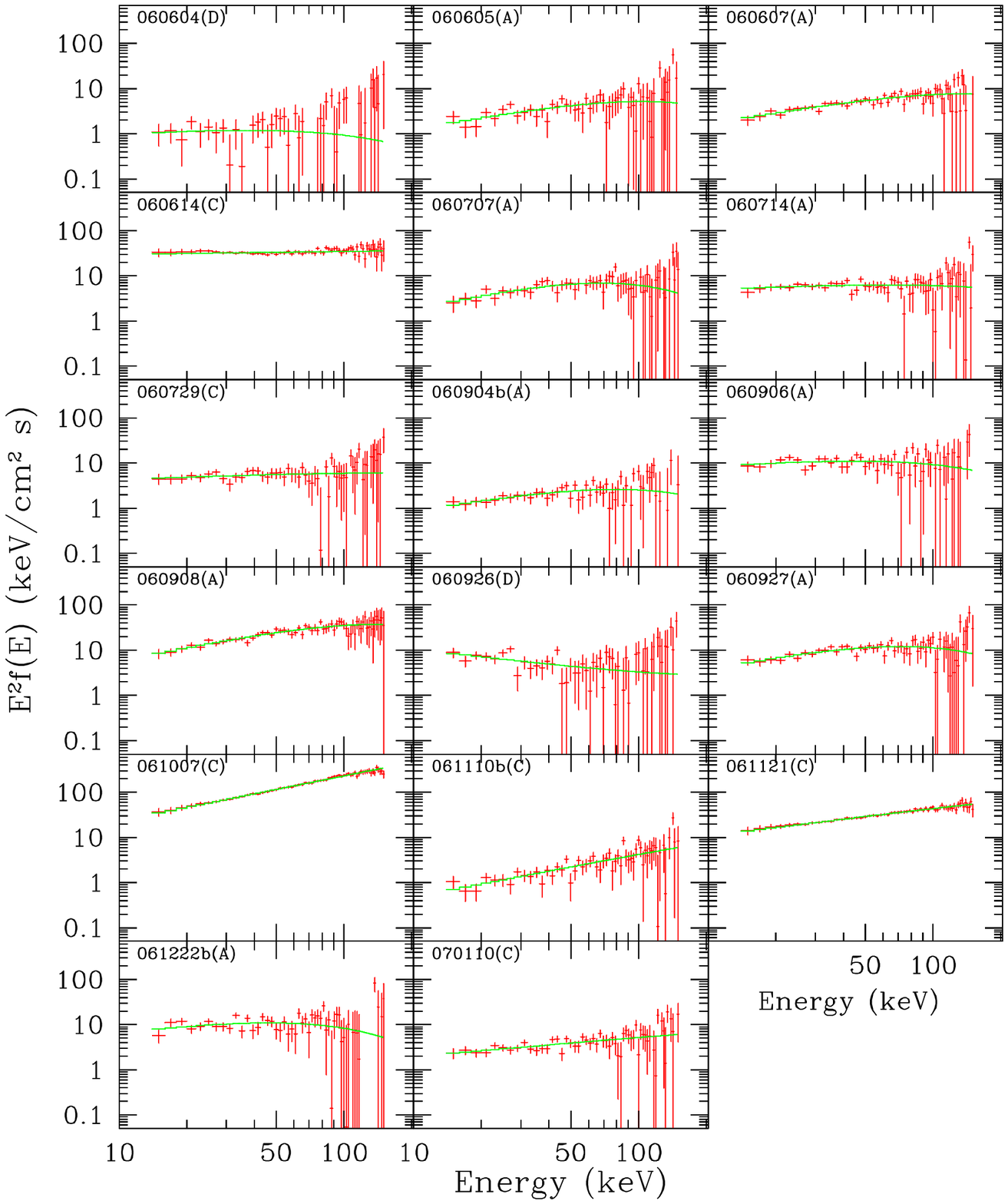,width=17.8cm,height=22cm}}
\caption{Continues Fig. \ref{spectra1}.}
\label{spectra3}
\end{figure*}

In this Appendix section we present the 47 \sw~ spectra (see Table \ref{PLfit}) 
analysed here. As described in \S 2, the 60 energy channels in the range 15--150 
keV of the BAT detector are used to deconvolve the spectra. The spectral fluxes
were averaged over the integration time, \tdur, determined for each burst from its
light curve (see \S 2). In Figs. \ref{spectra1}, \ref{spectra2}, and 
\ref{spectra3}, the CPL model was used to fit the observed spectra.

\end{document}